# Multi Agent Driven Data Mining For Knowledge Discovery in Cloud Computing

Vishal Jain[1] and Mahesh Kumar Madan[2]

[1]Research Scholar, Computer Science and Engineering Department Lingaya's University, Faridabad
[2]Professor and HOD, Computer Science and Engineering Department, Lingaya's University, Faridabad
[1]vishaljain83@ymail.com, [2]mahesh.madan@gmail.com

**ABSTRACT**
*Today, huge amount of data is available on the web. Now there is a need to convert that data in knowledge which can be useful for different purposes. This paper depicts the use of data mining process, OLAP with the combination of multi agent system to find the knowledge from data in cloud computing. For this, I am also trying to explain one case study of online shopping of one Bakery Shop. May be we can increase the sale of items by using the model, which I am trying to represent.*
**Keywords:** Data Mining, OLAP, Knowledge Discovery, Cloud Computing, Multi agent

## 1. Introduction

Cloud computing and data mining have become famous phenomena in the current application of information technology. With the changing trends and emerging of the new concept in the information technology sector, data mining and knowledge discovery have proved to be of significant importance. Data mining can be defined as the process of extracting data or information from a database which is not explicitly defined by the database and can be used to come up with generalized conclusions based on the trends obtained from the data. A database may be described as a collection of formerly structured data. (Song II – Yeol et al 15) Multi agents data mining may be defined as the use of various agents cooperatively interact with the environment to achieve a specified objective. Multi agents will always act on behalf of users and will coordinate, cooperate, negotiate and exchange data with each other. An agent would basically refer to a software agent, a robot or a human being (Wrembel & Koncilla 45) Knowledge discovery can be defined as the process of critically searching large collections of data with the aim of coming up with patterns that can be used to make generalized conclusions. These patterns are sometimes referred to as knowledge about the data. Cloud computing can be defined as the delivery of computing services in which shared resources, information and software's are provided over a network, for example, the information super highway. Cloud computing is normally provided over a web based service which hosts all the resources required (Bramer 36)

This paper will critically analyze the statement" Multi Agent driven data mining for knowledge discovery in cloud computing"

## 2. Data Mining

Data mining and knowledge discovery are very useful concepts especially when determining unknown phenomena. For example, a chain store that sells a variety of clothing can employ data mining skills to come up with some decisions. For instance, if it is observed that, at the beginning of the year more sweaters and jumpers are bought, and as the year progresses more light clothes are bought then the management can stock more jumpers and sweaters at the beginning of the year and ensure that they are sold at full price. (Zhou 55) In the above example, the database does no store the purchasing pattern, but it stores the records of items bought and at what time. For us to be able to establish the pattern, we need an agent which will observe the purchase pattern. On the other hand, the knowledge learnt from the above example is that more jumpers and sweaters are bought at the beginning of the year probably due to the winter season. Light clothes are bought during the mid year period may be due to the summer season.

## 3. Cloud Comuting
Cloud computing data mining can be very useful to cloud vendors. Cloud vendors can obtain useful information about their customers and use it to make valuable decision. In obtaining, information from the cloud computing data mining, the cloud vendors should not obtain data for the purposes of reselling. For example if a company a company that is interested in collecting the information that most users look for over the internet can employ the web usage mining technique. This technique will gather information about what most web users access on the internet, whether it is text or multimedia information. (Cooley 22) Web hosting is an example of a cloud computing in the sense that most of the data is stored by the web hosting company and not the individual or organization owning the website. Services can be obtained on a website by the help of a browser. (Masand & Spiliopoulou 65)

## 4. Case Study
### 4.1 Problem Definition
The following model illustrates how data mining can be used to obtain information about the ordering behaviors of customers, who places their orders online using our interactive website. The information obtained will be used to identify which months produces the highest orders and by how much.

### 4.2 Solution
I have employed the use of OLAP method to analyze the ordering trends of customers. When customers visit our site, for them to be able to order a product online they need to click on the link, for place an order. Various information is captured and stored in a database table. A customer is supposed to enter the following details on the order form. FIRSTNAME, LASTNAME, GENDER, MARITAL STATUS, LOCATION, TYPE OF CAKE, PAYMENT MODE, DATE and AGE.

| DATE | FNAME | LNAME | GENDER | MARITAL STATUS | LOCATION | TYPE OF CAKE | PAYMENT MODE | AGE |
|---|---|---|---|---|---|---|---|---|
| 12/2/2011 | LEANZS | LIGALE | MALE | SINGLE | NAIROBI | VANILA | MASTER CARD | 18 |
| 12/2/2011 | KEVIN | NGAIRA | MALE | SINGLE | KIAMBU | VANILA | MPESA | 22 |
| 13/2/2011 | MILCAH | ADEMA | FEMALE | MARRIED | KISII | MILKY | PAYPAL | 32 |
| 13/2/2011 | STEVE | LUBITA | MALE | SINGLE | NAIROBI | VANILA | CREDIT CARD | 21 |
| 28/3/2011 | EDWIN | MAFUNU | MALE | SINGLE | KITUI | WEDDING | MPESA | 28 |
| 28/3/2011 | KITEN | KOLO | MALE | SINGLE | KIAMBU | WEDDING | PAYPAL | 30 |
| 29/3/2011 | KEVO | POLOP | MALE | SINGLE | NAIROBI | WEDDING | MPESA | 32 |

Table1. "The table indicates an extract of some data stored in the orders table from the database."

From the above data, we have to extract data that can be used to sturdy the ordering behaviors of the customers which will help us come up with a conclusion that will enable us increase our sales. The above database does not provide explicit data on purchasing behaviors of customers hence we have to perform pre – data processing. After pre data processing, we will then load the information into a relational database and later represent the information in a multi dimensional array. The multidimensional array will help us analyze the data using data cube and the OLAP (on- line analytical processing) of the relational database. Pre- data processing (Bramer 40).

Pre- data processing will involve various activities including, data cleaning, client details identification, capturing the order date and the type of cake that has been ordered by various clients. Data cleaning involves removing unwanted data from the large collection of data. For example, from the above model, we do not need data about the mode of payment, first name, last name, age, gender and location, hence data cleaning will remove the unwanted details and only remain with the wanted fields. Data cleaning can be achieved by using specialized intelligent software agents, or by other software applications like SPSS, or by use of a structured query language that will execute the selected query statement (Witten et al 78)

The table2 indicates the data that will be stored in the relational database ready to perform the OLAP process.

| DATE | MARITAL STATUS | TYPE OF CAKE |
|---|---|---|
| 12/2/2011 | SINGLE | VANILA |
| 12/2/2011 | SINGLE | VANILA |
| 13/2/2011 | MARRIED | MILKY |
| 13/2/2011 | SINGLE | VANILA |
| 28/3/2011 | SINGLE | WEDDING |
| 28/3/2011 | SINGLE | WEDDING |
| 29/ 3/2011 | SINGLE | WEDDING |

Table2. "Data in Relational Database"

An OLAP process uses the multidimensional data to represent information. The above data has to be represented into a multidimensional array. The above three attributes, DATE, MARITAL STATUS and TYPE OF CAKE can be represented in the following three dimensional array.

| DATE | MARITAL STATUS | TYPE OF CAKE | COUNT |
|---|---|---|---|
| 12/2/2011 | SINGLE | VANILLA | 2 |
| 13/2/2011 | MARRIED | MILKY | 1 |
| 13/2/2011 | SINGLE | VANILLA | 1 |
| 28/3/2011 | SINGLE | WEDDING | 2 |
| 29/3/2011 | SINGLE | WEDDING | 1 |

Table3. "Ordered Data"

The table3 data is then converted into a data cube that will eventually allows us to perform the roll up, drill down, slicing and the dicing operations on the model. (Witten 79)

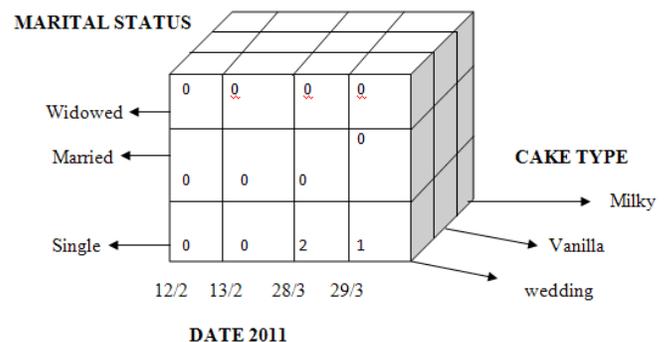

Figure1. "Iris cube that represents the data which can now be sliced and diced."

The Figure1 data cube represents all the possible aggregates of data which is as a result of a proper selection of a subset of the dimensions and summing up all the remaining solutions. In the above data cube, those combinations that have not been represented are indicated using a zero. The boxes that have been represented are indicated by the count number they indicate. From the above data cube we can conclude that on 28/3/2011, two clients who were single bought the wedding type cake.

In order to properly analyze the above data we need to split into individual cubical components that represent the data. We have to perform the slicing and dicing operation on the data. Slicing is the process of selecting a range of cells or a group of cells from the multi dimensional array through specifying a specific value. On the other hand data dicing refers to the process of defining a sub array from the entire multidimensional array. (Cios 48)

The tables (table4, table5, table6) below show the cross tabulation tables after the slicing and dicing operation has been done.

a). vanilla cake

|         | 12/2/2011 | 13/2/2011 | 28/2/2011 | 29/2/2011 |
|---------|-----------|-----------|-----------|-----------|
| single  | 2         | 1         | 0         | 0         |
| married | 0         | 0         | 0         | 0         |
| widowed | 0         | 0         | 0         | 0         |

**Table4. "Ordered Data of Vanilla Cake"**

b). Milky cake

|         | 12/2/2011 | 13/2/2011 | 28/2/2011 | 29/2/2011 |
|---------|-----------|-----------|-----------|-----------|
| Single  | 0         | 0         | 0         | 0         |
| married | 0         | 1         | 0         | 0         |
| widowed | 0         | 0         | 0         | 0         |

**Table5. "Ordered Data of Milk Cake"**

c). wedding cake

|         | 12/2/2011 | 13/2/2011 | 28/2/2011 | 29/2/2011 |
|---------|-----------|-----------|-----------|-----------|
| Single  | 0         | 0         | 2         | 1         |
| Married | 0         | 0         | 0         | 0         |
| widowed | 0         | 0         | 0         | 0         |

**Table6. "Ordered Data of Wedding Cake"**

After performing the data slicing and dicing, we have then to obtain a fact table that will give the final results which can now be used to make conclusions. The table7 indicates data which can be obtained from the above data cube.

|         | 12/2/2011 | 13/2/2011 | 28/3/2011 | 29/3/2011 | TOTAL |
|---------|-----------|-----------|-----------|-----------|-------|
| WEDDING |           |           | 2         | 1         | 3     |
| VANILLA | 2         | 1         |           |           | 3     |
| MILKY   |           | 1         |           |           | 1     |
| TOTAL   | 2         | 2         | 2         | 1         | 7     |

**Table7. "Extracted Data"**

We have been able to extract a table that shows how much cakes were sold on which date and the same time the types of cakes sold at what time of the year. The above data is a two dimensional representation from the three dimensional array. Other two dimensional data can still be obtained like what type of cake is mostly liked by a certain marital status group and at what time. The above data is not explicitly provided by the database. Viable conclusions could be made from the above data, we could obtain the knowledge that, (Cios 55)

1. Vanilla cakes, which are designed by our company for special moments with loved one is highly bought on the week just before the valentine day. Hence more vanilla cake could be baked and sold on that date at a maximum price.
2. Wedding cakes are highly bought towards the end of March. This could be probably attributed to the fact that most weddings are carried out in the month of April and many people prefer buying weeding cakes at this time in preparation of the weddings, hence more wedding cakes should be produced at this moment and sold at a maximum price.

Also if we want to determine which marital status group likes our cakes most, we would come up with the following table from the data cubes

|  | MARITAL STATUS | SINGLE | MARRIED | WDOWED | TOTAL |
|---|---|---|---|---|---|
| WEDDING |  | 3 | 0 | 0 | 3 |
| VANILLA |  | 3 | 0 | 0 | 3 |
| MILKY |  | 0 | 1 | 0 | 1 |
| **TOTAL** |  | 6 | 1 | 0 | 7 |

**Table8. "Final Result"**

From the table8, we can conclude that most of our cakes are largely bought by young single men. This will help our marketing team now that we have been able to determine our target audience. This can be implemented by ensuring that most of our advertisements and marketing activities target the young men.

## 5. Conclusion

From the above model, we have obtained knowledge that implemented multi agent data mining that discovered knowledge in cloud computing since our database is hosted on a cloud computing platform.